\documentclass[twocolumn]{webofc}
\usepackage[varg]{txfonts}   % Web of Conferences font
\begin{document}
\title{Decoherence of collective motion in warm
nuclei}

\author{\firstname{S.} \lastname{Frauendorf}\inst{1}\fnsep\thanks{\email{sfrauend@nd.edu}} \and
        \firstname{C. M.} \lastname{Petrache}\inst{2}\and
        \firstname{R.} \lastname{Schwengner}\inst{3}\and
        \firstname{K.} \lastname{Wimmer}\inst{4}
}

\institute{University Notre Dame, USA 
\and
           University Paris Sud \& CSNSM-CNR, France
\and
           Helmholtz Center Dresden Rossendorf, Germany
\and  The University of Tokyo, Japan           
          }

\abstract{%section{summary}  
 Collective states in cold nuclei are represented by a wave function that assigns coherent phases to the participating nucleons.   
 The degree of coherence decreases with excitation energy  above the yrast line because of coupling
  to the increasingly dense background of quasiparticle excitations. The consequences of decoherence are discussed, starting with 
   the well studied case of rotational damping. In addition to superdeformed bands,  a highly excited oblate band is presented as a new example of screening from rotational damping.  Suppression of pair correlation 
  leads to incoherent thermal M1 radiation, which appears as an exponential spike (LEMAR) at zero 
  energy in the $\gamma$ strength function of spherical nuclei.  In deformed nuclei a Scissors Resonance appears and LEMAR changes 
  to damped magnetic rotation, which is interpreted as partial restoration of coherence.          }
\maketitle
\section{Introduction}
\label{intro}
\begin{figure}[t]
\centering
\includegraphics[width=\linewidth,clip]{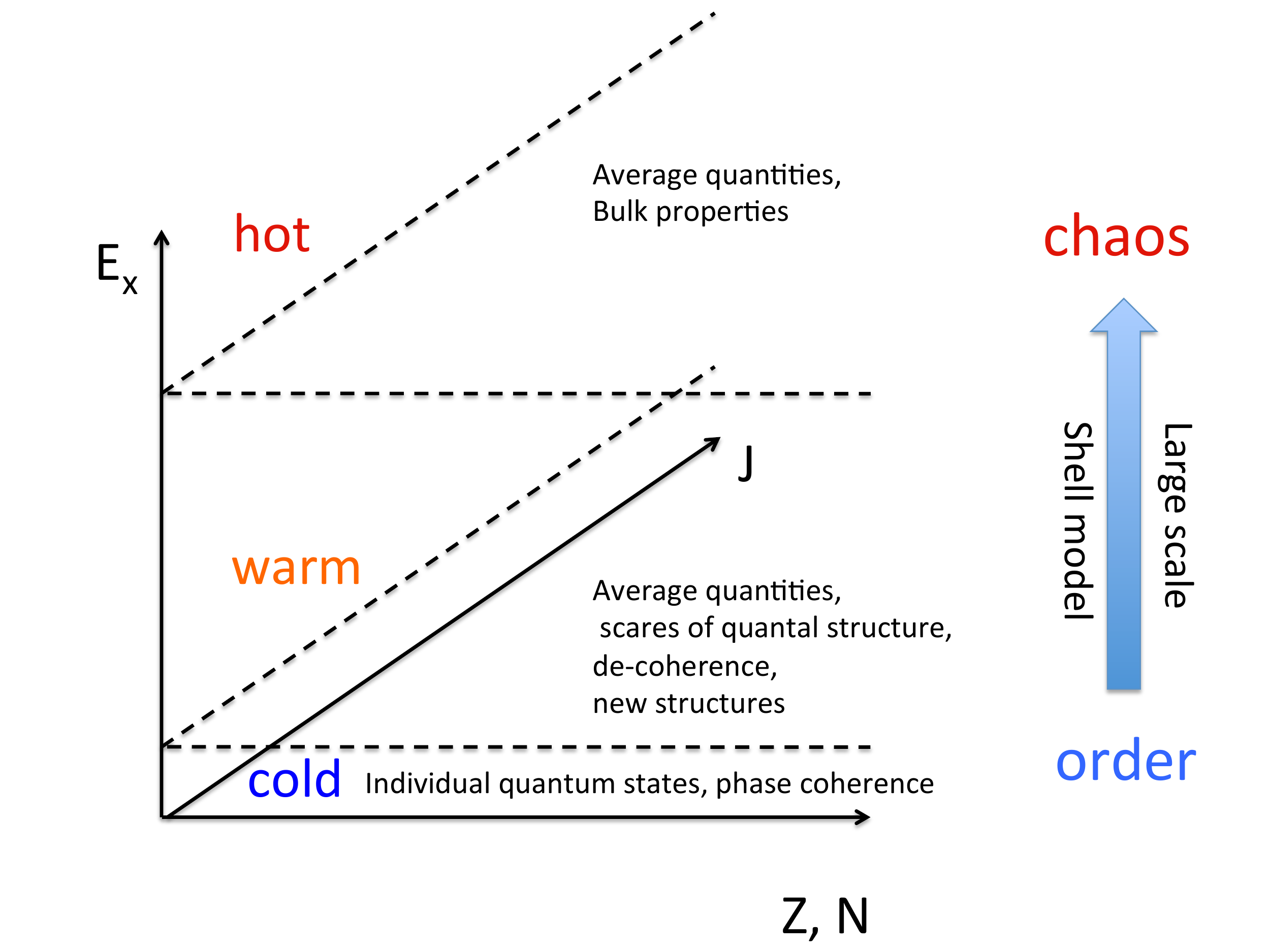}
\caption{The region of warm nuclei.}
\label{f:WarmNuclei}       % Give a unique label
\end{figure}

The study of various quantal phenomena in cold nuclei, i. e. near the yrast line, is the major thrust of the nuclear
structure community. The study of hot nuclei by means of heavy ion reactions is another active field. In this talk I address
the properties of "warm nuclei" which are excited by several MeV above the yrast line (see Fig. \ref{f:WarmNuclei}). My focus is the disappearance  of quantal 
phenomena characterizing cold nuclei, which is also referred to as the transition from order to chaos.
 I call this "decoherence of collective motion"  alluding to condensed matter physics where the 
analog loss of coherence is observed when the system's temperature increases (for example the disappearance of the 
pair condensate in superconductors/superfluids). Compared to macroscopic systems, nuclei are small enough for finding
"exact" solutions by numerical diagonalization of a model Hamiltonian in a finite basis. Such simulations of the nuclear dynamics can
be considered as the theoretician's experiment for warm nuclei.  The challenge is to distill simple features from the huge set of data
generated by the computer. 

The properties of warm nuclei are an important  input into network calculations for the stellar processes as well as for applications to 
nuclear engineering. One example are the strength functions for $\gamma$ emission below the neutron emission threshold.

\section{Collectivity vs. Coherence}
\label{sec:CollCoh}

\begin{figure}[h]
\centering
\includegraphics[width=0.95\linewidth]{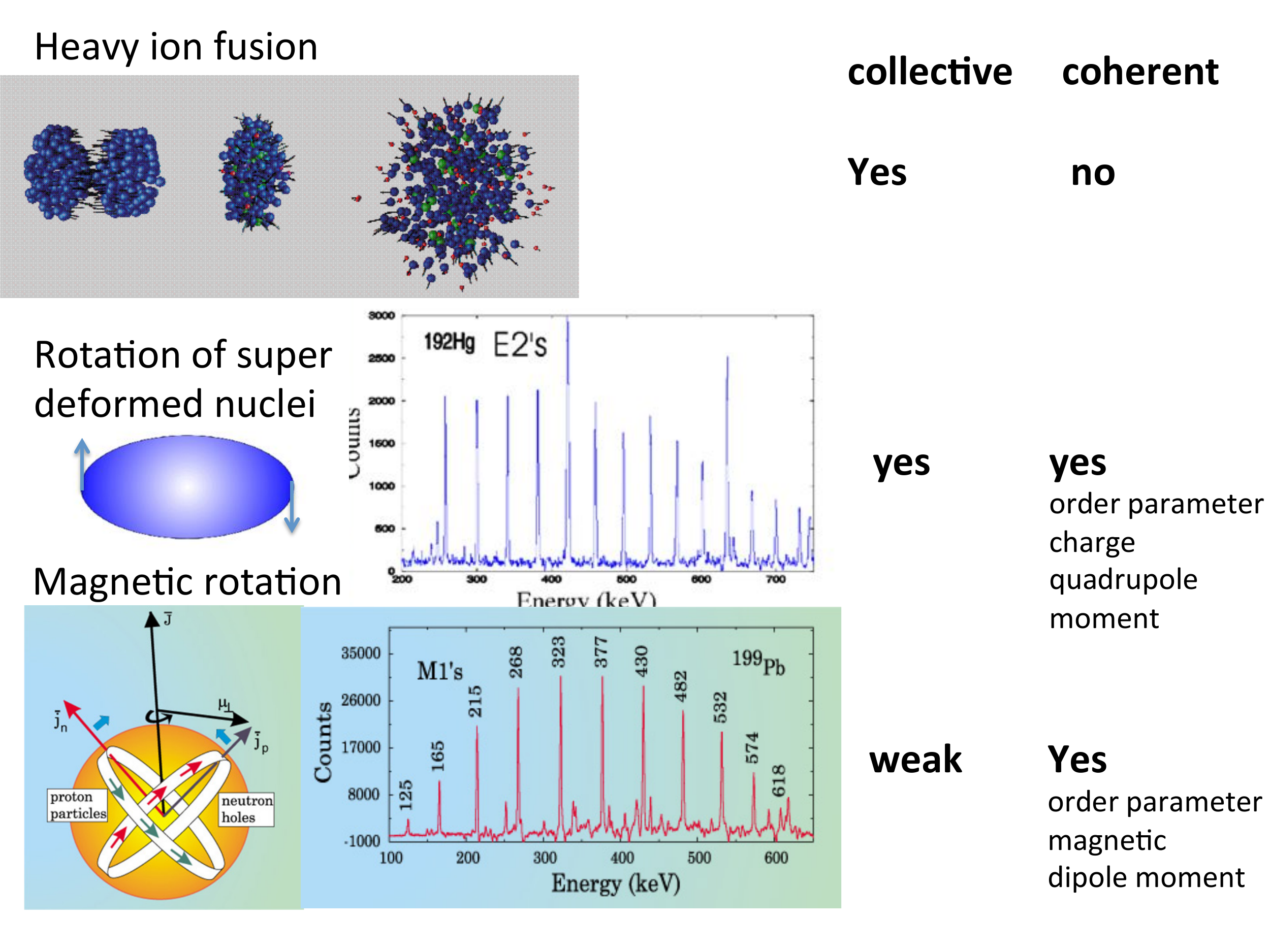}
\caption{Relation between collectivity and coherence.}
\label{f:ColCoh}       % Give a unique label
\end{figure}

Discussing cold nuclei, the termini "collective" or "collectivity" indicate the presence  
of a "collective wave function" that is carried by a large number a nucleons.
In the context of warm nuclei one needs sharpen terminology. 
A collective wave function implies the existence of definite phase relations between the constituents, which is called coherence in condensed matter physics. "Collective" implies that many constituents move in an organized way.
As illustrated in Fig. \ref{f:ColCoh}, collectivity does not necessarily imply coherence and vice versa. Heavy ion fusion is a collective
process, but not coherent. Magnetic rotation is carried by few nucleons, not very collective but coherent.  
The fixed phase relations between the nucleons lead to the emergence of the same  $I(I+1)$ energy spectrum that is characteristic for the 
highly collective rotation of superdeformed nuclei.    

The theory of superconductivity introduces the "coherence length", which  quantifies the degree of coherence. Loosely speaking,
it is the size of a Cooper pair. The quantal phenomena, as the Meissner effect, Josephson effect or flux quantization, 
emerge as  consequences of the existence of a wavefunction of the pair condensate. The coherence length is the 
minimal length on which the wave function lives. Its phase cannot increase more  than $2\pi$ within this interval. By
the Uncertainty Principle, this sets the upper limit of the momentum of the wave function and, as a consequence, 
it limits the super current density.

In analogy, Frauendorf \cite{CohAng01} introduced the coherence angle $\Delta \psi$  for nuclear rotation as the width of the overlap
between two mean field states of different orientation.  The complementary coherence length $\Delta I$ counts how many 
times $\Delta \psi$ fits into the angle range of $2\pi$ ( $\pi$ in case of ${\cal R}(\pi)=1$ symmetry), which  
 limits the angular momentum range of a rotational band by the Uncertainty Principle.
For the superdeformed bands  $\Delta \psi=5^\circ$ and 
 $\Delta I = \pi/\Delta \psi=36 $. For well deformed nuclei  $\Delta \psi=8^\circ$ and 
 $\Delta I =  \pi/\Delta \psi=22 $. For the weakly deformed magnetic bands  $\Delta \psi=25^\circ$ and
 $\Delta I = 2\pi/\Delta \psi=14 $. The angular momentum coherence length $\Delta I$ correlates well with the number of observed transitions
for the three types of rotational bands (see Fig. \ref{f:ColCoh}). Further details can be found in Refs.   
   \cite{CohAng16,CohAng18}. 
   
    There, the coherence of the axial $\beta$ vibration is discussed as well. 
   The coherence length $\Delta \beta$ decreases with the number of crossings between the single particle states as functions
   of the deformation $\beta$. 
   For the well deformed nuclei in the center of the rare earth region, $\Delta \beta$
   is too large to support the first vibrational excitation with one node. In the transitional region around $N=90$ it is short
   enough to support one node. This correlates with the experimental evidence for the respective absence or presence  
   of a collective $\beta$ vibrational excitation.

    In general, coherence becomes weaker with increasing excitation energy, which leads to phase transitions from ordered to
 more chaotic phases in extended systems. The coherence length increases with warming up nuclei, and the quantal collective 
 modes of cold nuclei dissolve. In the following the decoherence of the quantal rotational mode will be discussed in some detail.  

  \begin{figure}[h]
\centering
\includegraphics[width=0.95\linewidth]{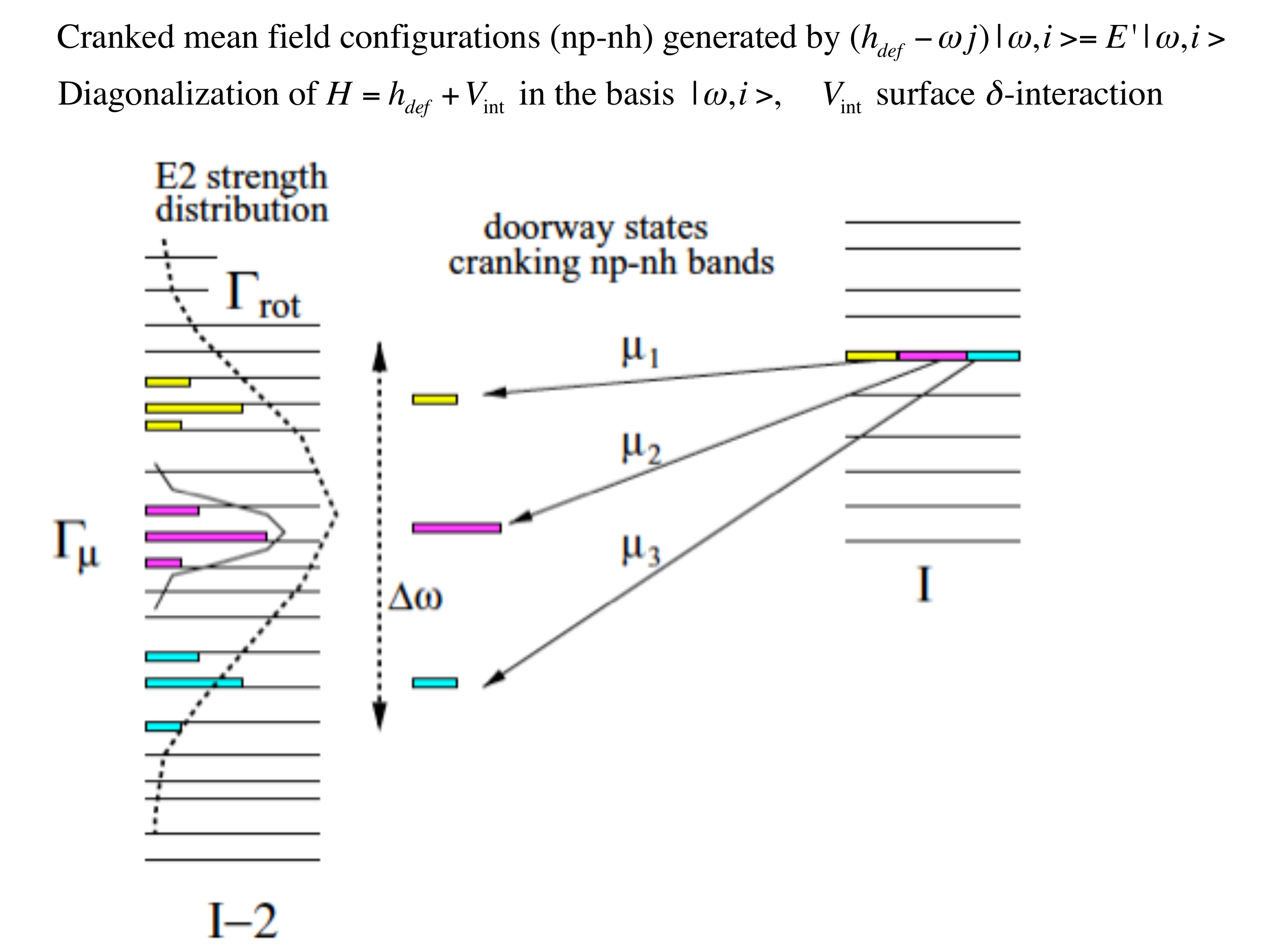}
\caption{Fragmentation of the E2-transition strength by band mixing. From Ref. \cite{RotDamp}. }
\label{f:BandMixing}       % Give a unique label
\end{figure}      
     
 \begin{figure}[h]
\centering
\includegraphics[width=0.9\linewidth]{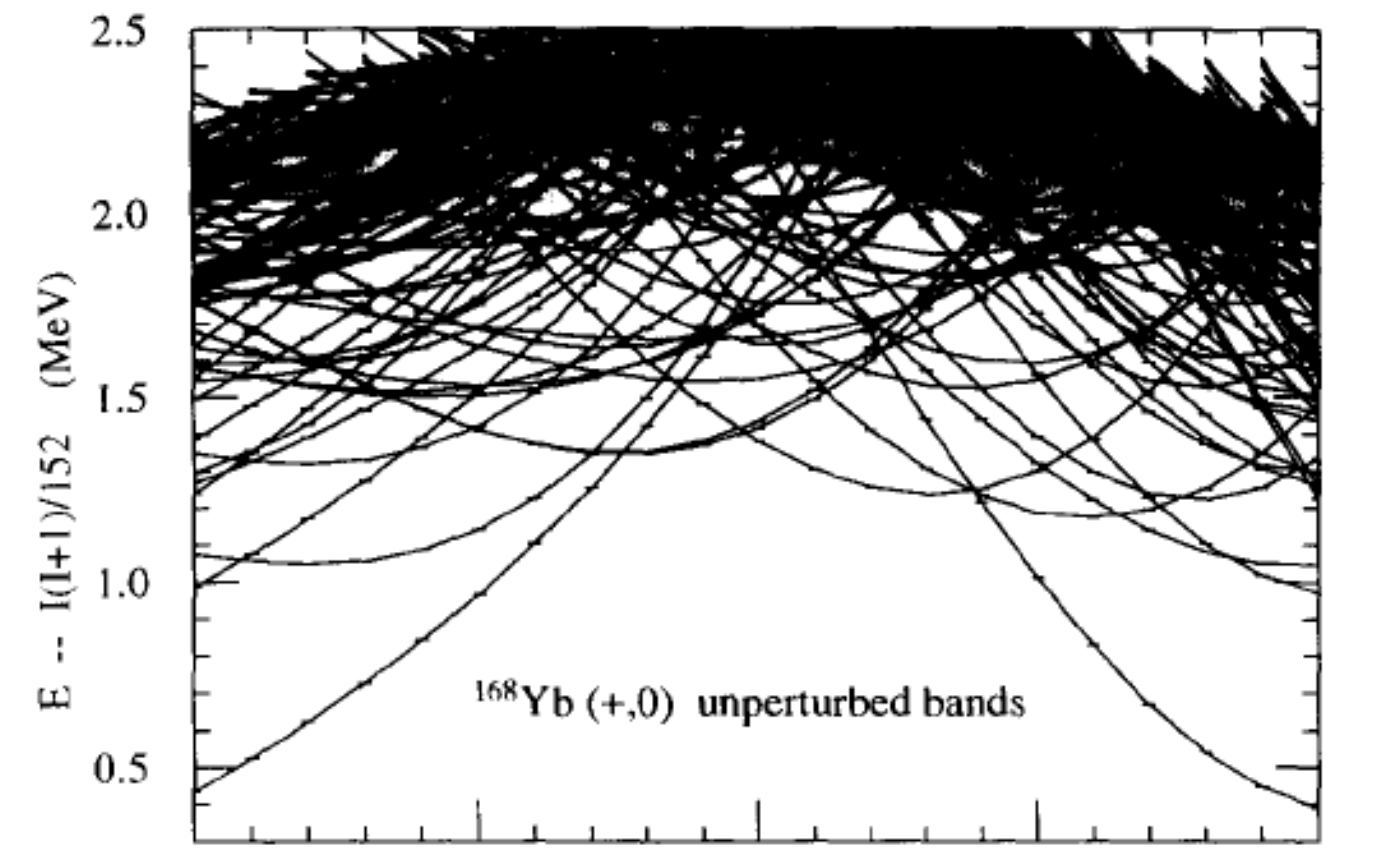}
\includegraphics[width=0.9\linewidth]{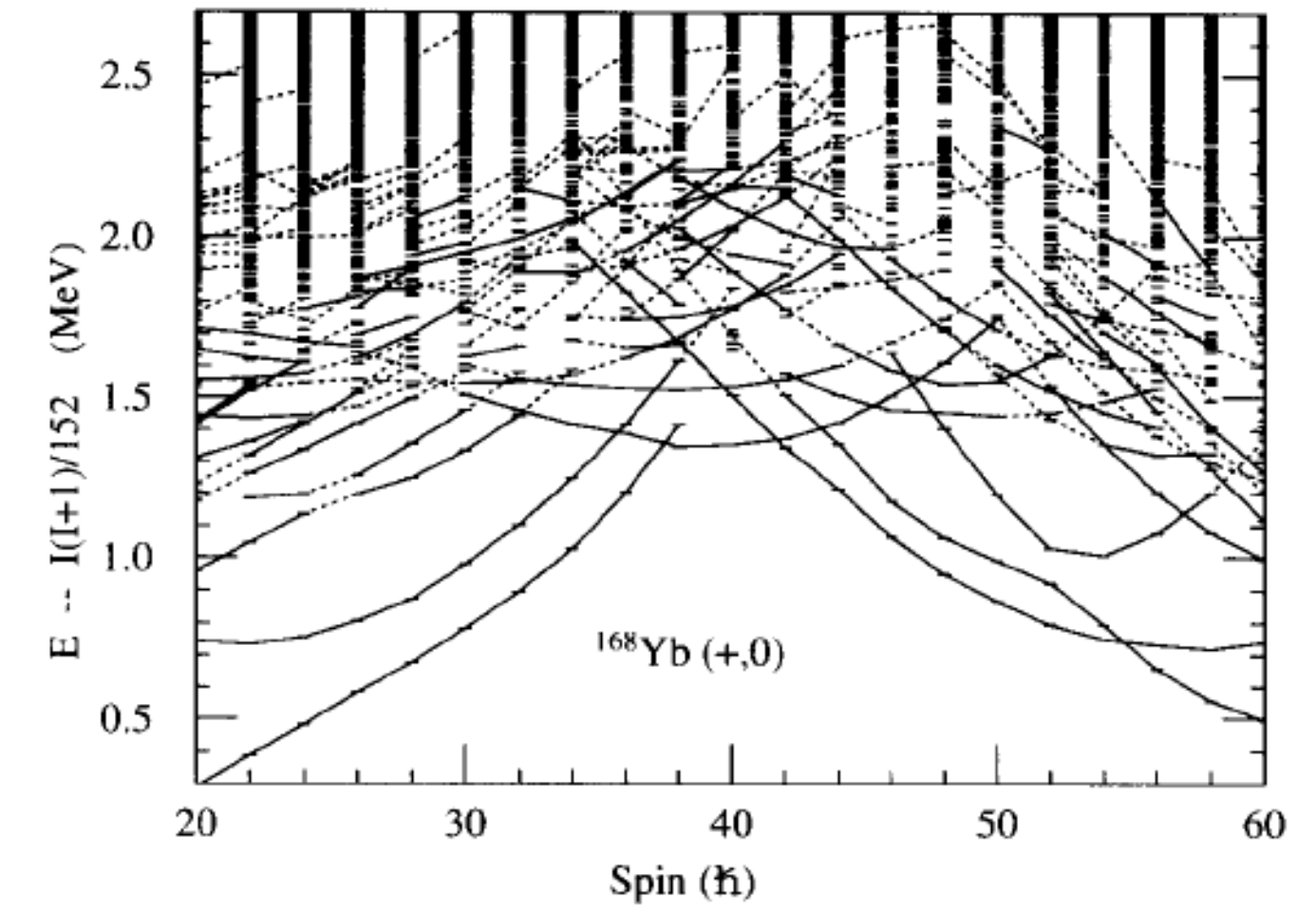}
\caption{Reduction of the E2 transition strength  by rotation damping. The short horizontal lines show the
energies of the states.   Upper panel (before mixing): the full lines trace individual
ph-configurations which represent discrete rotational bands. They are connected by transitions with large $B(E2,I\rightarrow I-2)$
values (assumed to be 1). Lower  panel (after mixing): the full lines connect mixed states for which  \mbox{
$B(E2,I\rightarrow I-2)>0.7$} and dotted lines for \mbox{$0.7>B(E2,I\rightarrow I-2)>0.5$}.
 From Ref. \cite{Matsuo}. }
\label{f:RotDeco}       % Give a unique label
\end{figure}      

\subsection{Rotational Damping}
\label{sec:RotDamp}

The coherence of rotational motion is manifest by the appearance of rotational bands, which are characterized by a smooth 
increase of the energy with angular momentum and strong intra-band E2 transitions. The latter are detected  by 
means of $\gamma - \gamma$ coincidence measurements,  which are THE TOOL to 
identify the members of a band. As such they probe the coherence length $\Delta I$ in  angular momentum.  

The Copenhagen/Milano/Kyoto collaboration carefully studied the progressive decoherence of quantal nuclear rotation, which they called "Rotational Damping". Recently  Leoni and Lopez-Martens presented the work in a concise review \cite{RotDamp}.
The theoretical analysis starts from the single particle configurations in a rotating deformed potential described by the
Cranked Shell Model. The various configurations, diabatically continued as functions of the rotational frequency,  well account for the discrete rotational bands observed in the region of about 1 MeV above the yrast line (e. g. see \cite{Frauendorf18}).

The particles are assumed to interact by a surface $\delta$-interaction, and the resulting  
Hamiltonian is numerically diagonalized.  Because of the exponential decrease of the level spacing with excitation energy, 
there appears strong band mixing, which is illustrated in Fig. \ref{f:BandMixing}.  The state $I$ is composed of three configurations
(yellow, pink, green). As each of which has a somewhat different moment of inertia a strong E2- transition $\mu_i$ arrives  
at  different energy at spin $I-2$.
The band mixing causes a fragmentation of each configuration by $\Gamma_\mu$, which generates a fragmentation of the strong E2 strength over states within $\Gamma_{rot}$, 
called  "rotational damping". 

The fragmentation dilutes the strong E2-transitions that serve as filter to thread a rotational band, which is illustrated in 
 Fig. \ref{f:RotDeco}. The intra band $B(E2, I\rightarrow I-2)$ values of the pure bands in the upper panel are  normalized to 1.
 The $B(E2 I\rightarrow, I-2)$ values  of the strongest transitions between the mixed states rapidly fall below 0.5. Together 
 with the finite resolution by the detectors this sets a limit to  identify discrete rotational bands.
 Still one finds a correlation between the unresolved E2 transitions in form of enhanced $\gamma - \gamma$ coincidence
 probability along ridges in the $E_\gamma-E_\gamma$ plane. The number $N_{ridge}$ of  resolved ridges can be taken as a measure
 of the coherence length $\Delta I=N_{ridge}\times 2$ of warm nuclei.  The typical number  $N_{ridge}=2,~3$ give
  $\Delta I =4,~6$ to be compared with  $\Delta I =22$  in cold rare earth nuclei.   

\begin{figure}[t]
\centering
\includegraphics[width=0.95\linewidth]{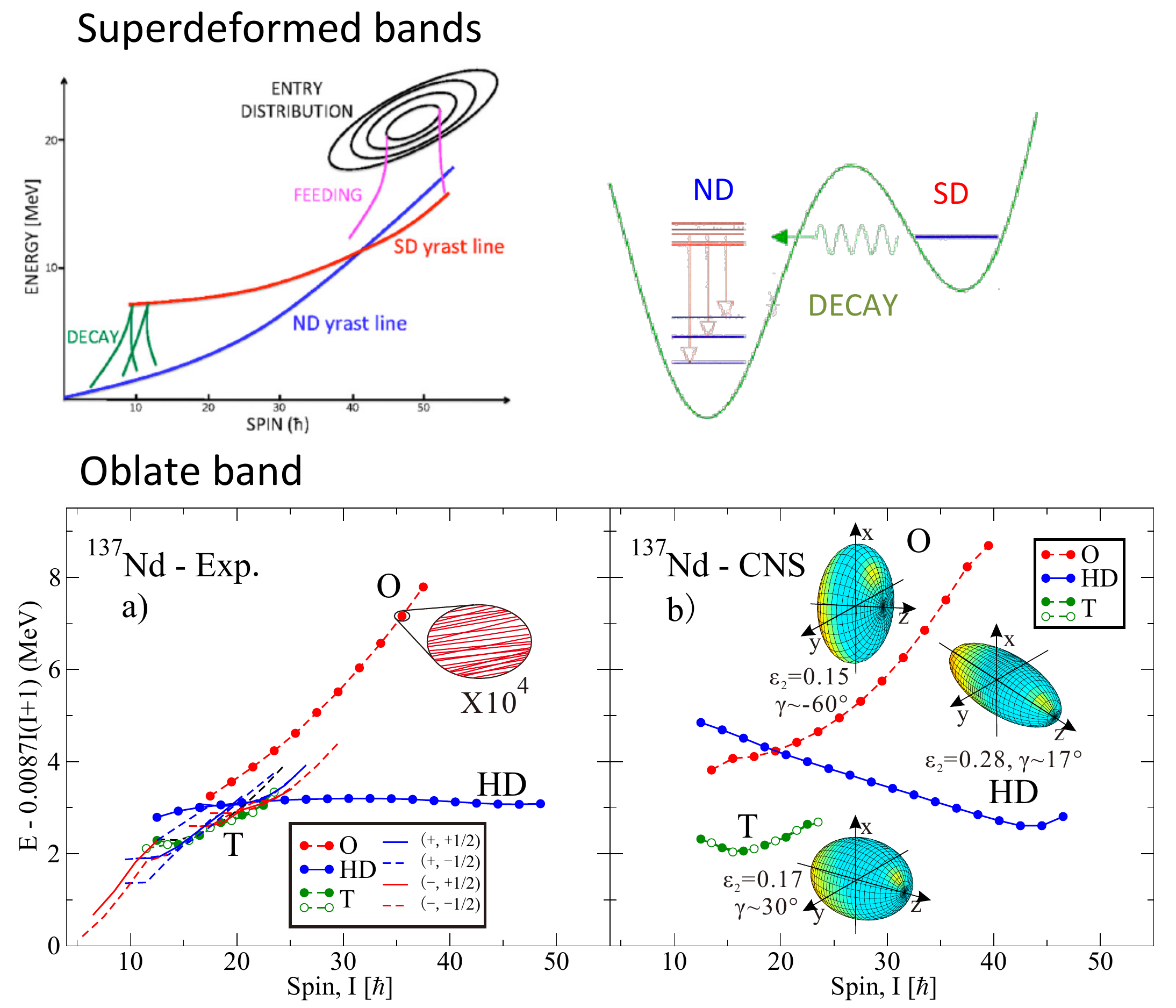}
\caption{Examples of bands screened from rotational damping. 
Upper panels: Superdeformed band separated by a potential well from the damped normal deformed states. 
From Ref. \cite{RotDamp}. Lower panels: Oblate band in $^{137}$Nd screened from the damped highly deformed states by
re-occupation of many orbitals. Left: experimental energies. Right: CNS calculations. From Ref. \cite{oblate}.}
\label{f:oblate}       % Give a unique label
\end{figure}

  \section{Islands of order in the sea of chaos}\label{sec:oblate}  
 Not all excited bands are rotationally damped. The upper panels of Fig. \ref{f:oblate} illustrate the well studied case of 
 superdeformed bands. They are populated at high spin, where they make up the yrast region. They dispose of their
 angular momentum by emitting E2 radiation, while becoming excited with respect to the normal deformed states.
   Around $I=10$ they are 5-8 MeV above yrast line, embedded in the dense environment of rotationally damped 
   normal deformed states. Their mixing with the background state is suppressed by a potential well in the deformation
   degrees of freedom. The tiny admixtures do not disturb the regular band pattern and their decay branches cannot 
   compete with the strong intra-band E2 transitions. At the lowest spins the coupling increases and the band de-excites
   via fragmented path ways.
     
The lower left panel of Fig. \ref{f:oblate} presents $^{137}$Nd as a new case of a regular band 
that extends  5 MeV above the yrast line well in to the 
damping region. The lower right panel shows calculations in the framework of the Cranked Nilsson Strutinsky (CNS) model, which well 
reproduce the experimental band structure.  There are three classes of states. For $I<20$ the yrast region consist of weekly deformed 
triaxial (T) bands, for $I>20$  it consist of highly deformed (HD) , slightly triaxial bands. The third class are oblate (O) configurations 
to which we assign the highly nonyrast band.  In contrast to the superdeformed bands the CNS calculation does not give a potential well
between the O and HA configurations. We attribute the suppression of the coupling to the substantial structural difference. Relative
to the $Z=50$, $N=82$ core the states are built on the lowest configurations  
\\HD: $[\pi(h_{11/2}^4(sdg)^6)\times\nu(h_{11/2}^{-4}h_{9/2}^2i_{13/2}^1(sdg)^{-8})]$
\\O:~~~~$[\pi(h_{11/2}^2(sdg)^8)\times\nu(h_{11/2}^{-1}(sdg)^{-4})]$. \\
 The O band acts like a funnel. Any feeding transition is followed by fast E2 transitions down to the region of triaxial bands, where O dissolves.

 \section{Low Energy Magnetic Radiation}\label{sec:LEMAR}
 Using their new analyzing technique of the $\gamma$ quasi continuum, the Oslo group discovered an enhancement  of the
 $\gamma$ strength function (GSF) below the transition energy of 2 MeV \cite{Guttormsen05}. 
 The observation came as a surprise, because the dipole excitations from the ground state do not show such an enhancement. 
 It was also met with skepticism, because it was on variance with the generally accepted Brink-Axel hypothesis, which states that
 GSF depends only on the transition energy $E_\gamma$ but not on the energy of the initial state.
 Later experiments demonstrated the systematic appearance of the enhancement (see e. g. Appendix II of Ref. \cite{Mitbo18}).

\begin{figure}[t]
\centering
\includegraphics[width=0.495\linewidth]{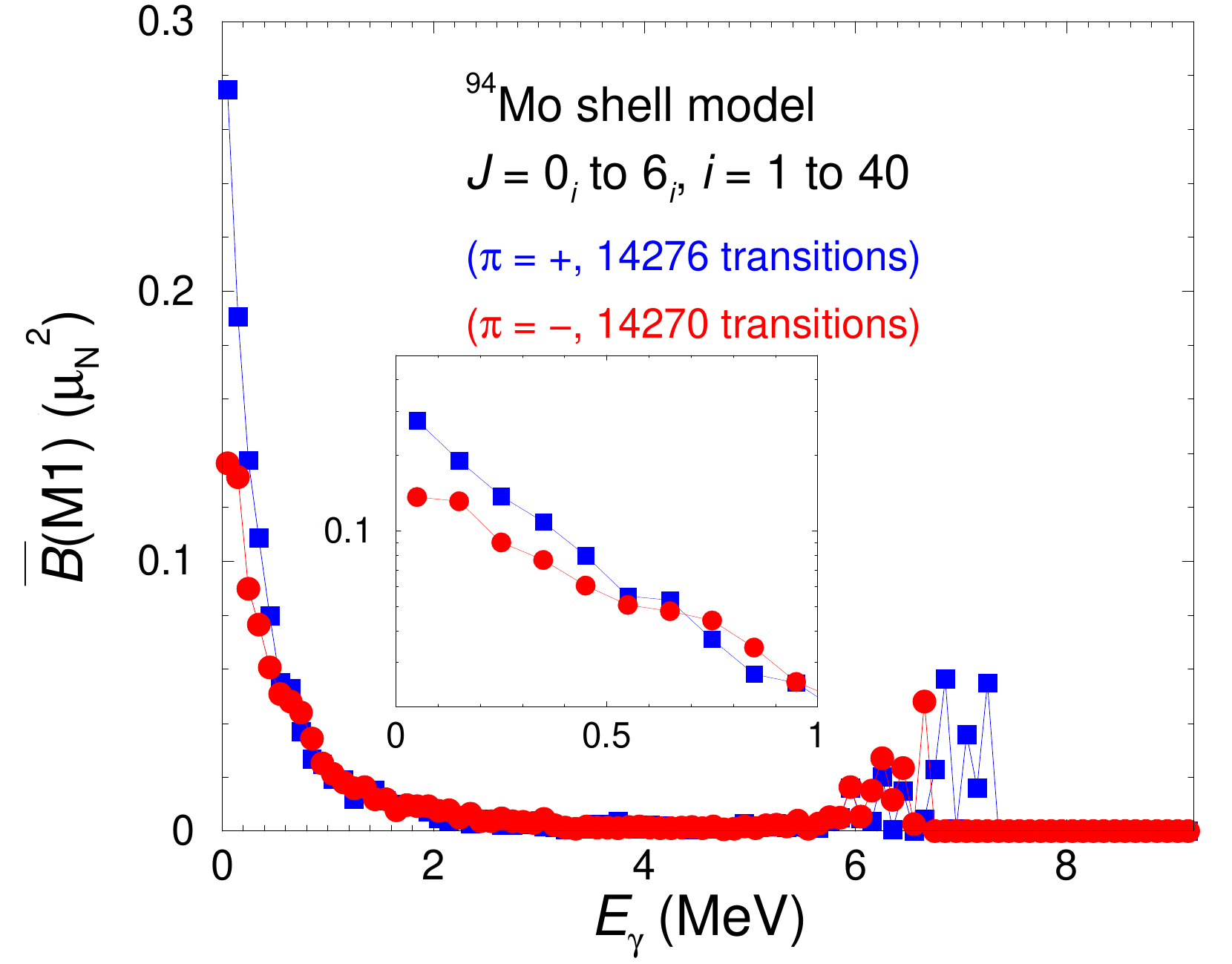}
\includegraphics[width=0.495\linewidth]{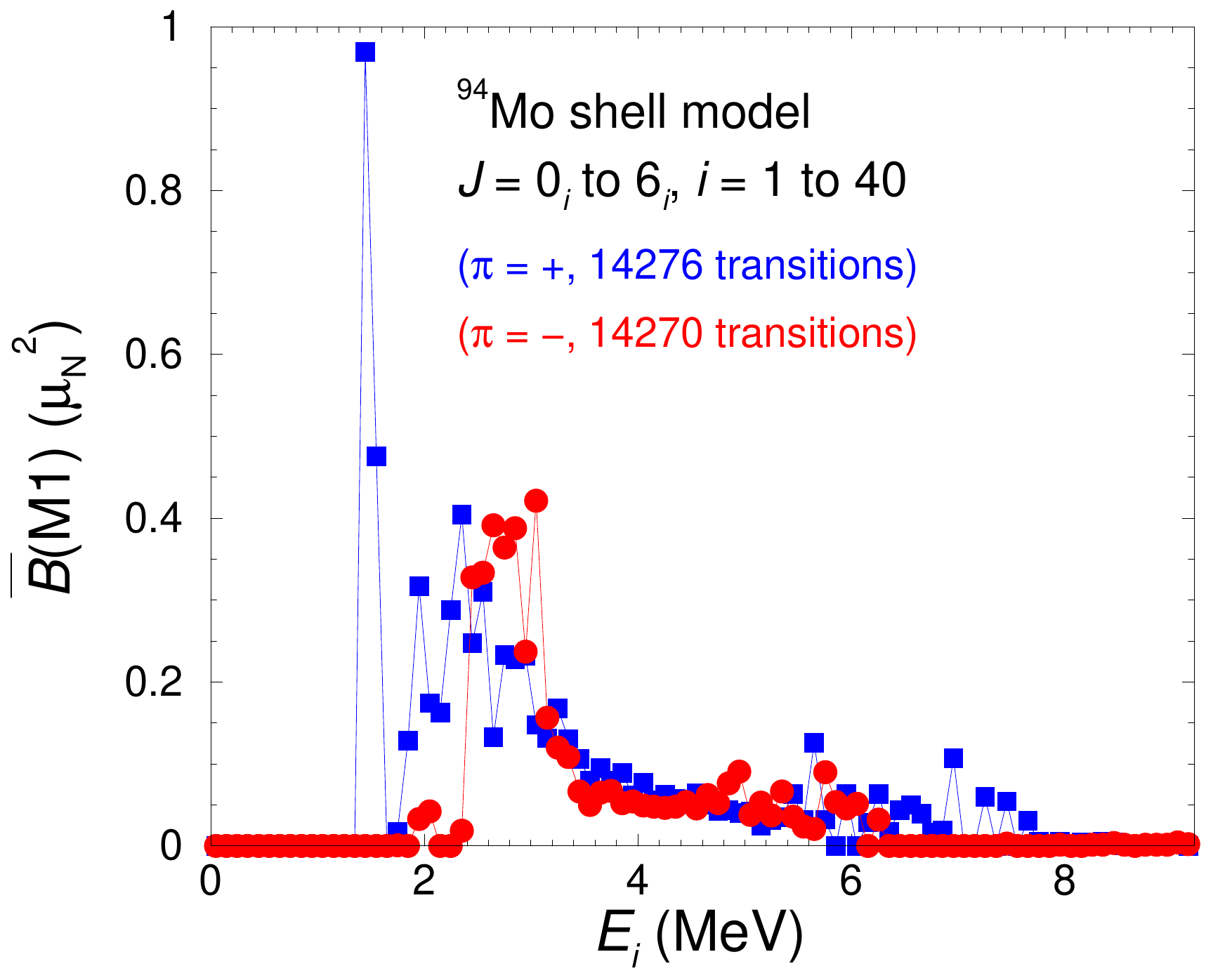}
\caption{(Color online) Left: Average $\overline B(M1)$ values in 100 keV
bins of $E_\gamma$ and averaged over all $E_i$ bins  for positive-parity (blue squares) and
negative-parity (red circles) states in $^{94}$Mo. The inset shows the
low-energy part in logarithmic scale. Right: Average $\overline B(M1)$ values in 100 keV of $E_I$
averaged over the $E_\gamma< 5~MeV$ bins.
From \cite{Schwengner13}.}     
\label{f:LEMAR94Mo}
\end{figure}

  Shell Model Calculations for $^{94}$Mo and neighbors 
 with a $Z=28$, $N=40$ core provided an explanation of the phenomenon \cite{Schwengner13}. The authors
 calculated the $B(M1)$ for all allowed transitions between the lowest 40 
 states of given angular momentum $I=0-6$. The large set of data was analyzed as follows (present version,
 the original is a special case). The values were collected into $[E_\gamma,E_{i}]$ bins of 100 keV width.
  Average $\overline{B}(M1)$ values  were obtained by dividing the cumulative values by the number of transition
  allowed for each bin and  taking the mean value over a range of the initial energies $E_i$ of the transitions. 
 As seen in Fig. \ref{f:LEMAR94Mo}, there is a strong enhancement of the average $\overline{B}(M1)$ values at
 $E_\gamma=0$ which falls off approximately exponentially with a constant $1/T_B$, where $T_B=0.49$ MeV. 
The  exponential decrease is retained by GSF, which was calculated by 
\begin{equation}
f_{M1}(E_\gamma)
= \frac{16\pi}{9 (\hbar c)^{3}}\overline{B}(M1,E_\gamma)\rho(E_i),
\end{equation}
where the level density  $\rho(E_i)$ was obtained from the number of initial states in the $E_i$ bins.
 The authors of Ref.   \cite{Schwengner13} associated the experimentally observed enhancement of the GSF below 2 MeV
 with the calculated spike  at $E_\gamma=0$, which they called 
  Low Energy Magnetic Radiation (LEMAR). The presence of exponential LEMAR spike in weakly deformed nuclei was confirmed by subsequent Shell Model studies along the same line for the Fe isotopes \cite{Schwengner17} and the nuclei in the
  fp-shell \cite{Mitbo18,Karampagia17,Sieja18} (and earlier work cited). 
 
   \begin{figure}[t]
\centering
\includegraphics[width=\linewidth]{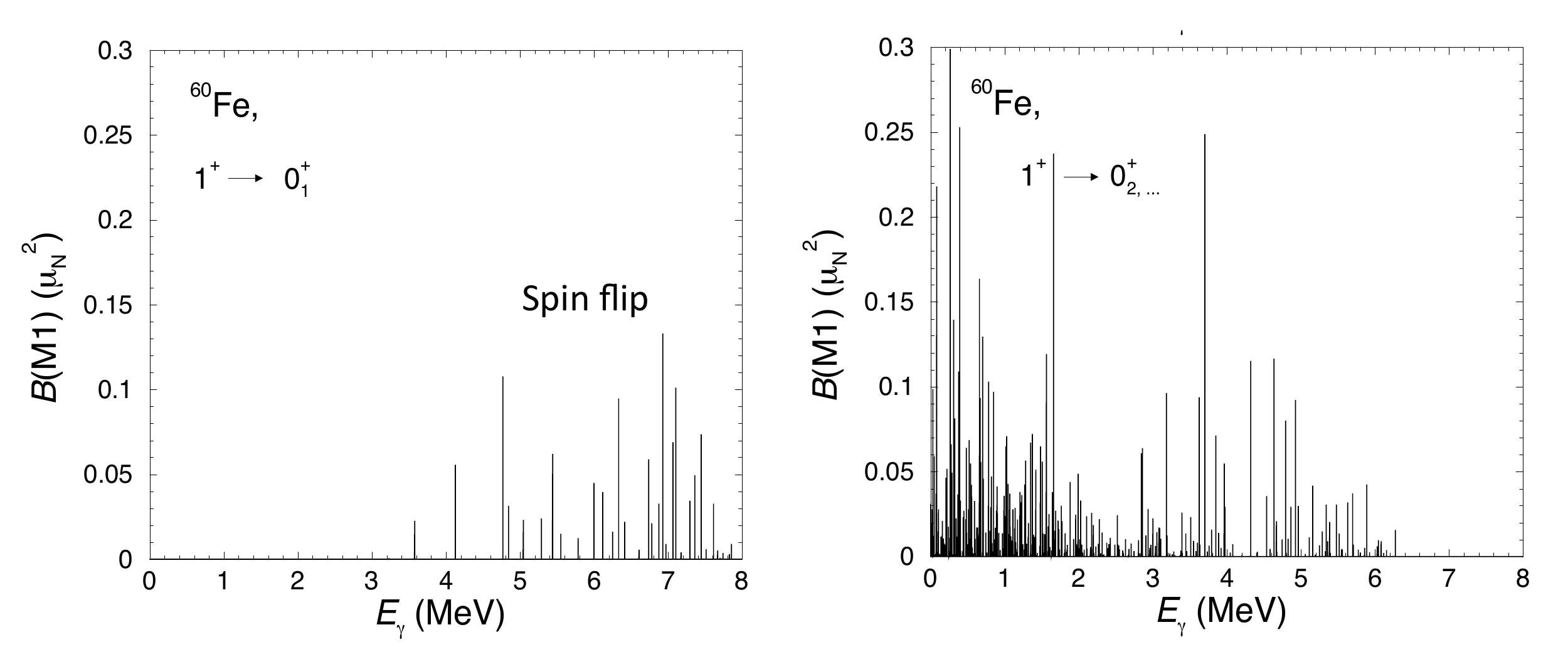}
\caption{ Reduced transition probabilities ${B}(M1)$ of all transitions ending in the ground state $0^+_1$
and the first excited state  $0^+_2$ of $^{60}$Fe. Generated from the Shell Model calculations of Ref. \cite{Schwengner17}.}
\label{f:M1transitions}       
\end{figure}

\begin{figure}[h]
\centering
\includegraphics[width=\linewidth]{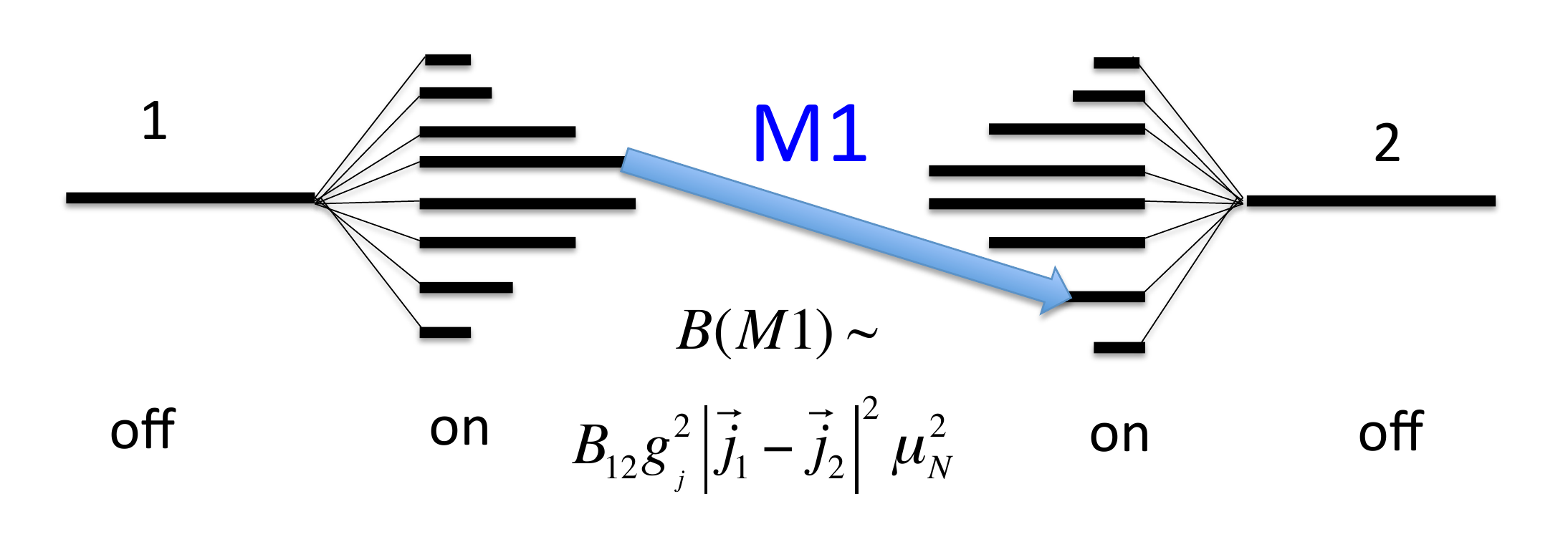}
\caption{ Two states 1, 2 within one and 
the same partition that are related by mutual re-alignment of the spins of the active high-j orbitals are
connected by a large M1 matrix element. Without residual interaction (off) the two states have the same energy $\rightarrow$ no radiation.
The residual interaction mixes the states in a chaotic way (on), which
generates energy differences between them and fragments the M1 strength $\rightarrow$ radiation.}
\label{f:splitting}       
\end{figure}

 The low-energy transitions appear only when they end in excited states (see right panel of Fig. \ref{f:LEMAR94Mo} and Fig. \ref{f:M1transitions}).   
The analysis of the initial and final states showed that transitions involve a reorientation of high-j orbitals within one and the same
partition of protons and neutrons (e. g.  in $^{94}$Mo: $\pi g_{9/2}^2\times \nu(d_{5/2}^1g_{7/2}^1)$, and  $\pi(p_{1/2}^{-1} g_{9/2}^3)
\times \nu(d_{5/2}^2g_{7/2}^1g_{9/2}^{-1})$, and more). The reorientation of a high-j orbital has a large transitional magnetic moment
of $\vec \mu_{12}=g_j(\vec j_2-\vec j_1)$ which is reflected by the $B(M1)$ value. The transitional magnetic moment is particularly large
when the reorientation involves protons and neutrons such that they enhance each other (like in the case of magnetic rotation the
simultaneous reorientation high-j protons and high-j neutron holes \cite{CohAng01,CohAng16,CohAng18}). As illustrated in Fig. \ref{f:splitting}, without residual interaction the states in one partition that differ by 
the mutual orientation of the occupied j-orbital 
have all the same energy, and no radiative transitions appear. 
The  mixing of these states by the residual interaction
generates a splitting of the complex eigenstates, 
among which the low-energy radiative transitions appear.  A calculation for $^{94}$Mo with artificially reduced scale of the residual interaction  
by a factor of 2 gives $T_B=0.24$ MeV which is one half of $T_B=0.49$ MeV found with the full interaction strength.
This demonstrates that the width of the LEMAR spike is determined by the mixing.

The fragmentation of the two states reduces their contribution to $B(M1)\sim B_{12}\left| g_j(\vec j_2-\vec j_1)\right|^2$, where $B_{12}$ is 
a stochastic factor that measures the probability of the pair [1 2] contributing to the transition between the two mixed  states.   
The stochastic factor can be estimated assuming random mixing. The average probability of one of the basis states to be present in
the complex state is
$\propto 1/N$ with $N$ being the total number of basis states contributing to the mixed state. In the chaotic regime
the residual interaction is much larger than the level distance $d$. Therefore $N\propto1/d$ and 
$B_{12}\propto 1/(d_id_f)\propto \rho_i\rho_f$. At the energies of interest, the level densities $\rho_i$ and $\rho_f$  
are well reproduced by the "constant temperature" expression 
$\rho\propto \exp[E^*/T_L]$, which gives   $B_{12}\propto \rho_i^2\exp[-E_\gamma/T_L]$.  In fact, the Shell Model calculations of Ref.
  \cite{Schwengner13,Schwengner17} are consistent with an exponential fall-off of the GSF given by $T_B=T_L$, where $T_L$ is taken from the exponential increase of the level density in the same calculation (see Ref.\cite{Sanibel16} for details).   

Zelevinsky and co-workers \cite{Zelevinsky} discussed the relation of chaotic mixing of the basis states and the statistical concept 
of an ensemble at a temperature $T$. From this perspective, one may associate  $T_B$ and $T_L$ with the ensemble 
temperature $T$, and the appearance of the LEMAR spike may interpreted as follows.   In the ground state the nucleons on the high-j 
orbitals are frozen by the pair correlations which  couple them to zero spin pairs. To generate  an M1 transition, one has to break such a pair
which costs $2\Delta$ of energy. Warming up the nucleus destroys the pair correlations (decoherence). Now the nucleons on the high-j
orbitals are free to re-orient. Agitated by the warm environment they re-orient emitting radiation with characteristic 
 probability   $\propto \exp[-E_\gamma/T]$ per  gamma quant.  Thus, the LEMAR spike represents
incoherent thermal radiation of an ensemble of uncoupled magnetic dipoles generated by the nucleons on the high-j orbitals.      

LEMAR  boosts the rates of $(n,\gamma)$ reactions. We continued our study of nuclides around $^{132}$Sn (details see \cite{CGS15})
 which are a bottle neck in the r-process of element
synthesis in violent stellar events \cite{132SnTBP}. A LEMAR spike with a  GSF comparable to the Mo isotopes
 was found for all nuclides with at least 3 particles/holes with respect to the $Z=50$ and $N=82$ core. Its existence
 north-east of  $^{132}$Sn is at variance with the conjecture \cite{Schwengner13,CGS15} that it appears only in regions 
 where Magnetic Rotation is predicted \cite{CGS15}.
 
\section{Deformation}\label{sec:SR}
With $N$ and $Z$ moving away from closed shells deformation sets in, which represents a correlation between nucleons. 
The impact on the M1 GSF was studied in Ref. \cite{Schwengner17} for the $^{60,64,68}$Fe isotopes by means of 
Shell Model calculations using $Z=20$ and $N=28$ as a core. 
The calculations account for  the increase of 
quadrupole collectivity in the considered isotopic chain. For $^{60,68}$Fe
the respective experimental $B(E2,2^+_1\rightarrow 0^+_1)=190,~ 450 (eb)^2$
and  the calculated    $B(E2,2^+_1\rightarrow 0^+_1)=368,~ 406 (eb)^2$.
The values are consistent with  the transitional character of the ground state bands of all considered isotopes.

 \begin{figure}[t]
\centering
\includegraphics[width=\linewidth]{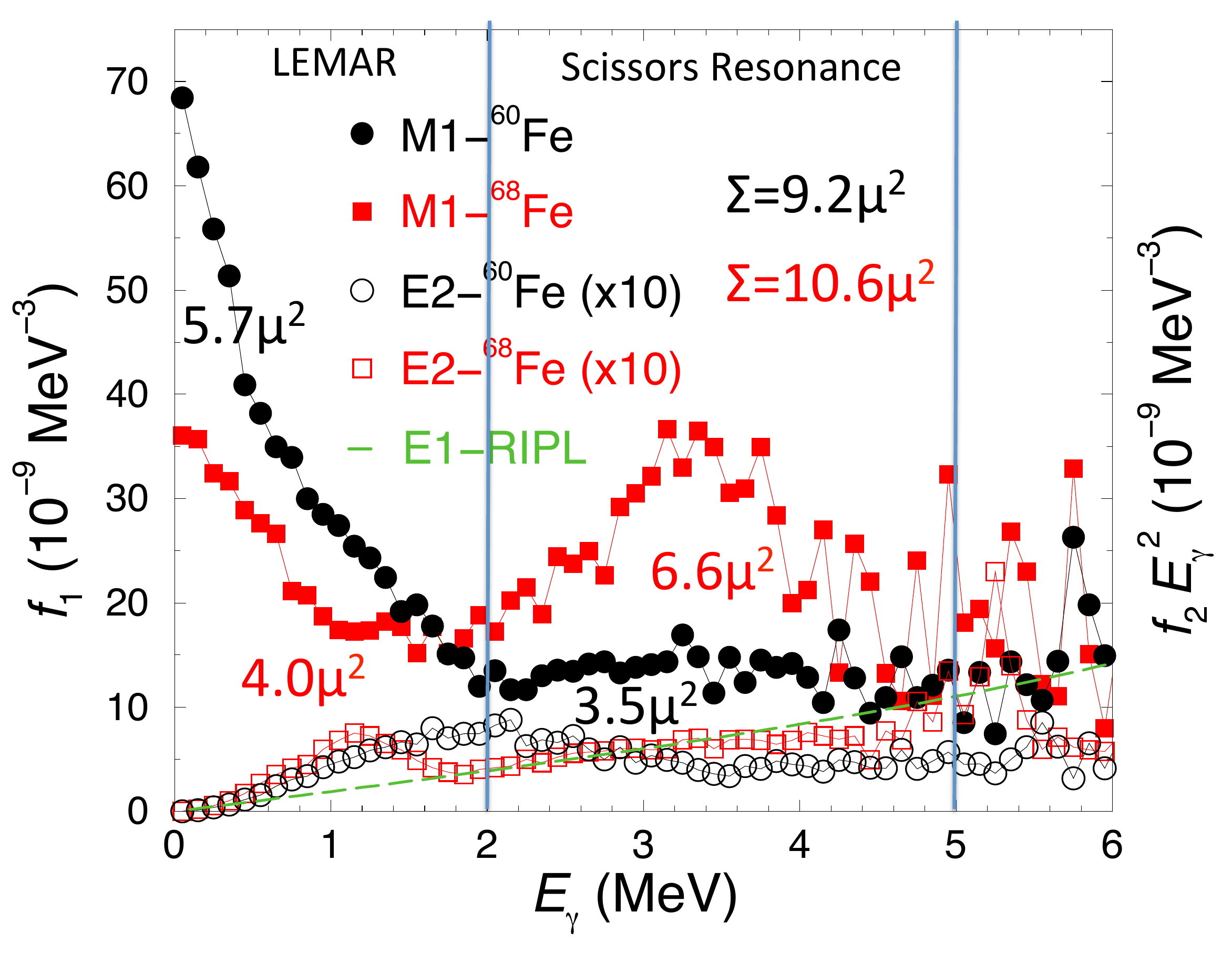}
 \caption{\label{f:f1Fe} (Color online) Calculated M1 strength functions ($f_1$)  and E2 strength functions $\left(f_2E_\gamma^2\right)$  in 
 $^{60,68}$Fe. Note the factor of 10 for  E2. The numbers quote the integrated strength (\ref{eq:IntStr}) where
 vertical lines indicate the respective integration regions. 
 For $^{68}$Fe the sum of the M1 strength from the ground state to all 1$^+$ states is 
 $\sum B(M1,0^+_1\rightarrow 1^+)=1.7\mu^2$. Adapted from Ref. \cite{Schwengner17}. }    
\end{figure}      

 Fig. \ref{f:f1Fe} shows the GSF for $^{60,68}$Fe. 
 The shape changes   from monomodal (only the LEMAR spike)  at $N$ = 34, four neutrons
above the closed shell, to bimodal at $N$ = 42, the middle of the $fpg$ shell.
    We interpret the bump in the
range from about 2 to 5 MeV as the Scissors Resonance (SR) built on excited states. 
The integrated  strength 
\begin{equation}\label{eq:IntStr}
\Sigma \overline B(M1)=\frac{9 (\hbar c)^{3}}{16\pi }\int_{E_{\gamma 1}}^{E_{\gamma 2}} f1(E_\gamma)d E_\gamma
\end{equation}
deduced from the strength functions 
below 5 MeV varies by only 8\% at most from an average of 9.80 $\mu^2_N$.
For the near spherical $^{60}$Fe it is mainly located in the LEMAR spike. For the deformed  $^{68}$Fe,
more than one half is shifted to the SR. 

Taking GSF for $^{64}$Fe into consideration, the authors of Ref. \cite{Schwengner17} concluded that the bimodal 
structure gradually develops with increasing $N$
by shifting  strength  from  the LEMAR spike to the SR while the sum stays nearly constant. 
The gradual development of the SR and the shift of the integrated strength when entering a shell has been confirmed
experimentally for odd-A Sm isotopes \cite{Naqvi19}. 
Systematic calculations using the full pf shell as configuration space
  for the Ge isotopes \cite{GeTBP} and the Ga isotopes \cite{Mitbo18} show that near the closed shells $N=28$ and 50 only
  the LEMAR spike exists. Going into the open shell from below and above the SR develops, first as a shoulder, and in the middle, where the deformation has a maximum, it becomes a distinct resonance.

 \begin{figure}[t]
\begin{center}
\includegraphics[width=\linewidth]{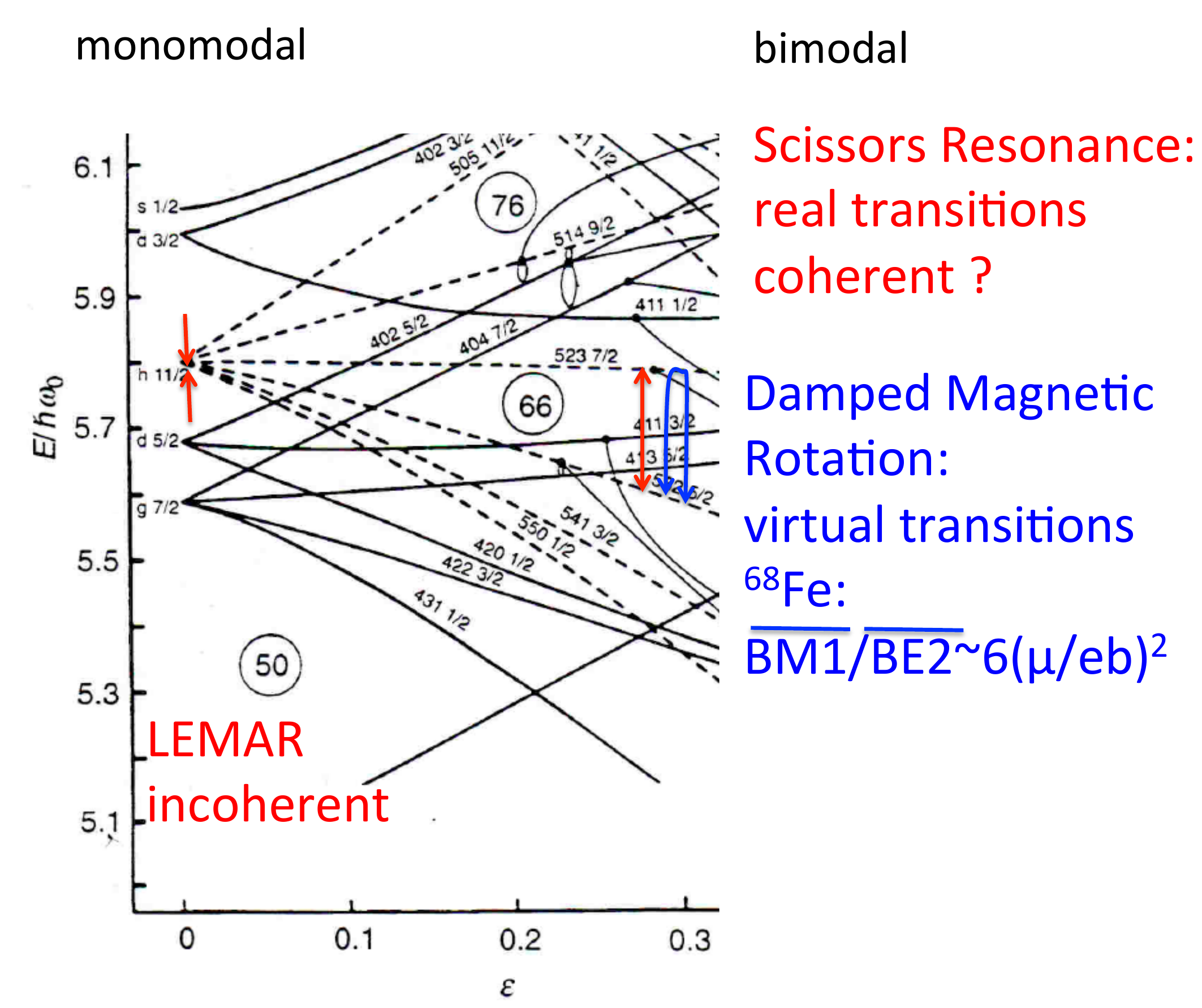}
\end{center}
 \caption{\label{f:Nilsson} (Color online) Transitions that generate the LEMAR spike and the Scissors Resonance. The energy
 scale is $\hbar \omega_0$=7.7 MeV for $A=$153.}
 \end{figure}
 \begin{figure}[h]
\begin{center}
\includegraphics[width=\linewidth]{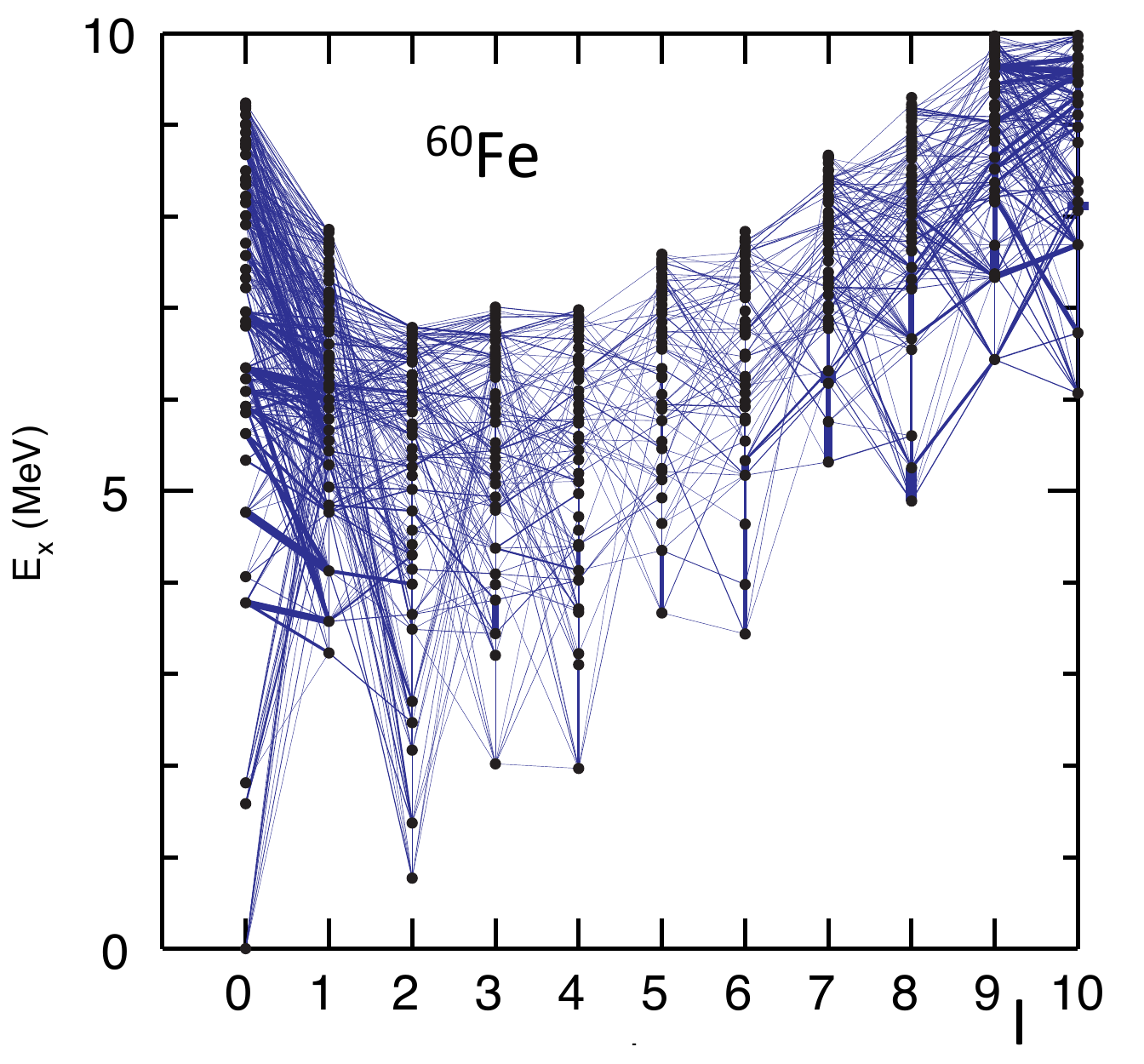}
%\end{center}
 %\caption{\label{f:spiderweb60} (Color online) Transitions with $B(M1)> 0.1\mu^2_N$
%(blue lines) between
%the calculated positive-parity states (black dots) in $^{60}$Fe. The
%widths of the lines are proportional to the $B(M1)$ values.}
% \end{figure}
% \begin{figure}[t]
%\begin{center}
\includegraphics[width=\linewidth]{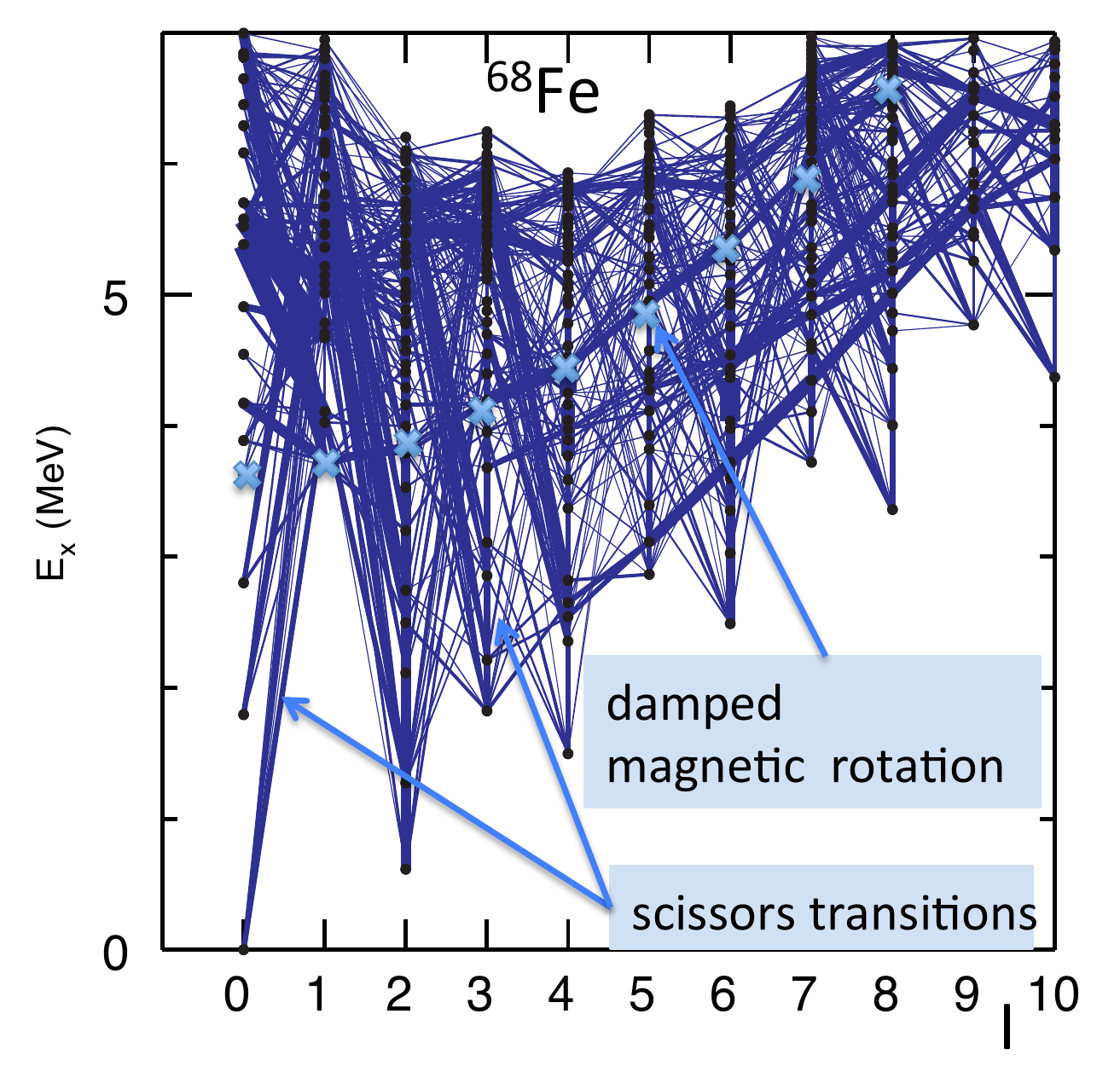}
\end{center}
 \caption{\label{f:spiderweb} (Color online) Transitions with $B(M1) > 0.1\mu_N^2$
 (blue lines) between
the calculated positive-parity states (black dots) in $^{60,68}$Fe. The
widths of the lines are proportional to the $B(M1)$  values.}
 \end{figure}

Fig. \ref{f:Nilsson} illustrates how the development of the bimodal LEMAR-SR structure can be explained by the 
onset of quadrupole deformation. 
 The mechanism causing large
$B(M1)$ values  is reorientation of the high-j single-particle angular momenta. Without deformation this 
occurs  between the various configurations of the nucleons in an incompletely filled j-shell ($h_{11/2}$ in the figure), 
which generates the  LEMAR spike as  discussed above. 
The magnetic (m) substates of the high-j multiplet split with the onset of deformation.  Reorientation 
occurs in two different ways. First, there are the real transitions between  the m-substates. Particle-hole
excitations of this type are known to generate the SR \cite{ham84}. As seen in Fig.  \ref{f:Nilsson},
the splitting between the m-states corresponds to the position of the SR for $\varepsilon=0.3$.  
Second, there are the virtual excitations from the occupied to the empty m-states. These are known
generating the rotational motion of deformed nuclei. Along 
a rotational band the angular momentum increases by a coherent gradual alignment of the spins 
of the nucleons  in the deformed orbitals, which is described by the 
virtual particle-hole excitations between the m-states. 
 
With increasing deformation, the
LEMAR spike changes from incoherent thermal-like radiation to partially coherent rotational radiation.
The ratio $B(M1)/B(E2) \sim 6(\mu/eb)^2$ is characteristic  for magnetic rotation \cite{CohAng01,CohAng16,CohAng18}.
As shown in Fig. \ref{f:spiderweb}, the M1 transitions between the states in $^{60}$Fe are chaotic. 
For $^{68}$Fe they acquire a certain 
preference along the curves   with $E_\gamma=I/{\cal J}$, which 
is indicated by sequence of  blue crosses in Fig. \ref{f:spiderweb}.  The moment of inertia ${\cal J} \approx 14$ MeV$^{-1}$
is close to the rigid body value  ${\cal J}_{rig} \approx 17$ MeV$^{-1}$. 
The preference is seen as the fuzzy threads, which signal the
onset of rotational coherence in form of damped magnetic rotation.  
The second preference are the strong transitions $I\rightarrow I\pm 1$ with an energy $E_\gamma \approx 3 MeV$, which 
generate the SR. It is not clear if their presence indicates vibrational coherence, which the concept of a collective Scissors Mode
implies. The may represent incoherent radiation from an ensemble magnetic dipoles, each of which oscillates independently
 in the deformed nuclear potential.

 S. F. acknowledges support by the DOE Grant DE-FG02-95ER4093.


\begin{thebibliography}{}


\bibitem{CohAng01} S. Frauendorf, Rev. Mod. Phys.  \textbf{73} 463 (2001) 
\bibitem{CohAng16} S. Frauendorf, AIP Conf. Proc \textbf{1753}, 030001 (2016)
\bibitem{CohAng18} S. Frauendorf, Phys. Scr.  \textbf{93}, 043003 (2018)
\bibitem{RotDamp} S. Leoni, A. Lopez-Martens, Phys. Scr. \textbf{91},  063009 (2016)
\bibitem{Frauendorf18} S. Frauendorf, Phys. Scr. \textbf{93}  043003, (2018)
\bibitem{Matsuo} M. Matsuo {\em et al.}, Nucl. Phys. A \textbf{617}, 1 (1997)
\bibitem{oblate} C. M. Petrache {\em et al.}, Phys. Rev. C {\bf 99}, 041301 (2019)
\bibitem{Guttormsen05} M. Guttormsen {\em et al.}, Phys. Rev. C {\bf 71}  044307 (2005)
\bibitem{Mitbo18} J. E. Mitb\o \, {\em et al.}, Phys. Rev. C {\bf 98}, 064321 (2018)
\bibitem{Schwengner13} R. Schwengner {\em et al.},  Phys. Rev. Lett. {\bf 111}  232504 (2013) 
\bibitem{Schwengner17} R. Schwengner {\em et al.}, Phys. Rev. Lett. {\bf 118}, 092502 (2017)
\bibitem{Karampagia17} S. Karampagia {\em et al.} Phys. Rev. C {\bf 95}, 024322 (2017)
\bibitem{Sieja18} K. Sieja, Phys. Rev. C {\bf  98}, 064312 (2018)
\bibitem{Sanibel16} S. Frauendorf {\em et al.} \textit{Proc. Int. Conf. on Fission and Properties of Neutron-rich Nuclei, 2016, Sanibel Island, Florida} World Science Pub. Singapore 2017 37-46; arXiv  1907.00398
\bibitem{Zelevinsky} V. Zelevinsky, Ann. Rev. Nucl. Part. Sci.  {\bf 46}, 237 (1996); V. Zelevinsky {\em et al.}, Phys. Rep. {\bf 391},  311 (2004)
\bibitem{CGS15} S. Frauendorf {\em et al.}, EPJ Web of Conferences {\bf 93} 04002 (2015), arXiv 1907.02641
\bibitem{132SnTBP} S. Frauendorf, K. Wimmer, to be published
\bibitem{Naqvi19} F. Naqvi {\em et al.}, Phys. Rev. C {\bf 99}, 054331 (2019)
\bibitem{GeTBP} S. Frauendorf, R.Schwengner,  to be published
%\bibitem{Raman} S.~Raman, C. W. Nestor JR., and P.~Tikaanen,  Atomic Data and Nuclear Data Tables, {\bf 78},  128, (2001) 
\bibitem{ham84} I. Hamamoto and S.  \AA berg, Phys. Lett. B {\bf 145}, 164 (1984)
\end{thebibliography}
\end{document}